\documentclass{article}
\usepackage{arxiv}

\usepackage[utf8]{inputenc} % allow utf-8 input
\usepackage[T1]{fontenc}    % use 8-bit T1 fonts
\usepackage{hyperref}       % hyperlinks
\usepackage{url}            % simple URL typesetting
\usepackage{booktabs}       % professional-quality tables
\usepackage{amsfonts}       % blackboard math symbols
\usepackage{nicefrac}       % compact symbols for 1/2, etc.
\usepackage{microtype}      % microtypography

\usepackage{graphicx}
\usepackage{subfig}
\graphicspath{ {./Figures/} }

\usepackage{listings}
\title{ABIDES: Towards High-Fidelity Market Simulation for AI Research}

\author{
  David Byrd \\
  School of Interactive Computing\\
  Georgia Institute of Technology \\
  Atlanta, GA 30308 \\
  \texttt{db@gatech.edu} \\
  \And
  Maria Hybinette\\
  Department of Computer Science\\
  The University of Georgia\\
  Athens, GA 30303 \\
  \texttt{maria@cs.uga.edu}\\
  \And
  Tucker Hybinette Balch \\
  School of Interactive Computing\\
  Georgia Institute of Technology \\
  Atlanta, GA 30308 \\
  \texttt{tucker@cc.gatech.edu}
}

\date{}

\begin{document}

\maketitle

\begin{abstract}
 We introduce ABIDES, an Agent-Based Interactive Discrete Event Simulation environment. ABIDES is designed from the ground up to support AI agent research in market applications. While simulations are certainly available within trading firms for their own internal use, there  are no  broadly available high-fidelity market simulation environments.  We hope that the availability of such a platform will facilitate AI research in this important area. ABIDES currently enables the simulation of tens of thousands of trading agents interacting with an exchange agent to facilitate transactions.  It supports  configurable pairwise network latencies between each individual agent as well as the exchange. Our simulator's message-based design is modeled after NASDAQ's published equity trading protocols ITCH and OUCH. We introduce the design of the simulator and illustrate its use and configuration with sample code, validating the environment with example trading scenarios.  The utility of ABIDES is illustrated through experiments to develop a market impact model.   We close with discussion of future experimental problems it can be used to explore, such as the development of ML-based trading algorithms.
\end{abstract}

\section{Background}
We have developed ABIDES, an agent-based interactive discrete event simulation, to facilitate the creation, deployment, and study of strategic agents in a highly configurable market environment.  We were inspired by Daniel Freidman's view that simulation provides a powerful tool to analyze individual participant behavior as well as overall market outcomes that emerge from the interaction of the individual agents.   In Freidman's review of empirical approaches to the analysis of continuous double auction (CDA) markets such as NASDAQ and the New York Stock Exchange, he outlines the strengths and weaknesses of three major approaches:
\begin{enumerate}
    \item Field studies of actual operating markets,
    \item Laboratory studies of small controlled markets,
    \item Computer simulation of markets.
\end{enumerate}
Freidman concludes that field studies are clearly relevant, but do not provide experimental access to all relevant information;  laboratory studies improve control and observation, but are of necessity small and expensive; and computer simulations feature perfect control and observation.  However ``trader's strategies are not endogenously chosen, but rather must be specified exogenously'' \cite{freidman1993double}.

As Freidman observed, market simulations provide an attractive platform for research in equity trading questions. This has led to the development of a number of simulation platforms such as those on which X. Wang and Wellman \cite{wang2017spoofing} and J. Wang et al. \cite{wang2017stockyard} have reported their results. We developed ABIDES as a fresh implementation to incorporate lessons learned from the deployment of prior platforms.  With ABIDES, we aim to address Freidman's primary concern regarding computerized market simulations -- that strategies must be exogenously specified -- with a platform enabling powerful \emph{learning} agents to easily participate in a realistically structured market via a common framework. We believe this is necessary to properly investigate the behavior and impact of intelligent agents interacting in a complex market environment. 

\begin{figure*}[h]
    \centering
    \includegraphics[width=6in]{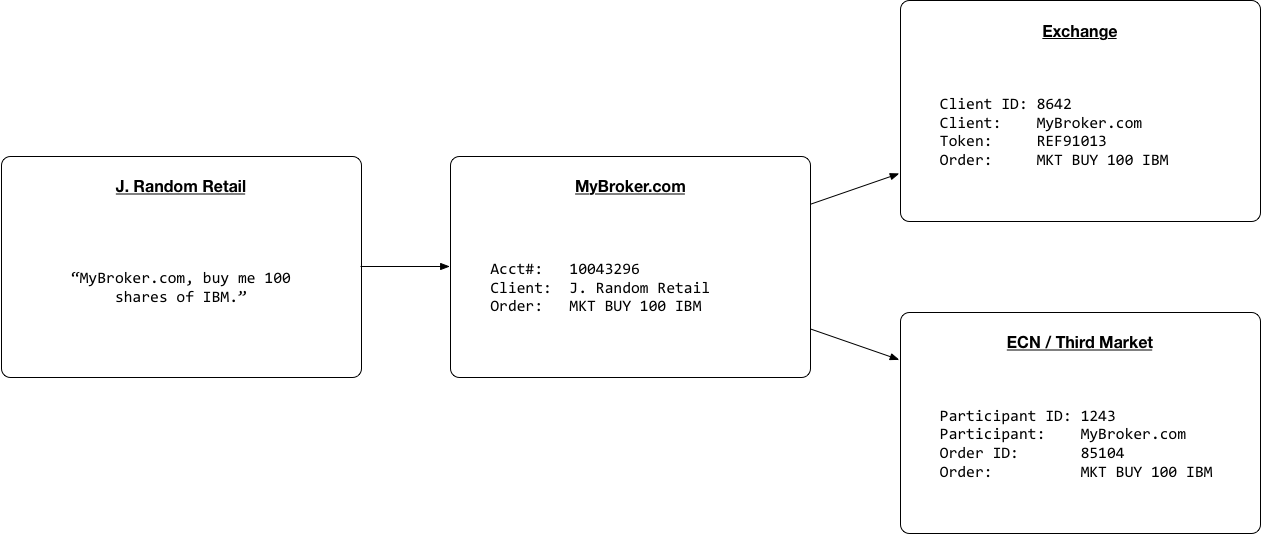}
    \caption{Simulation allows agent-identifiable data which is lost in the flow of real-world orders.}
    \label{fig:order_flow}
\end{figure*}

ABIDES is intended to be a curated, collaborative open-source project that provides researchers with \emph{tools} that support the rapid prototyping and evaluation of complex market agents.  With it, we hope to to further empower researchers of financial markets to undertake studies which would be difficult or impossible in the field, due to the absence of fine-grained data identifiable to individual traders (see Figure \ref{fig:order_flow}), a lack of knowledge concerning participant motivation, and an inability to run controlled ``what if'' studies against particular historical dates.

We acknowledge existing high-quality academically-targeted multi-agent market simulators such as that used by X. Wang and Wellman. In a recent study they used their simulation platform to study spoofing agents in a market environment populated by zero intelligence (ZI) and heuristic belief learning (HBL) traders  \cite{wang2017spoofing}.  Their approach analyzes the results from an empirical game-theoretic view \cite{wellman2006methods}.  We believe ABIDES makes a complementary contribution through its experimental focus on the ``market physics'' of the real world including:
\begin{itemize}
    \item Support for continuous double-auction trading at the same nanosecond time resolution as real markets such as NASDAQ;
    \item Ability to simulate specific dates in market history with gated access to historical data;
    \item Variable electronic network latency and agent computation delays;
    \item Requirement that all agents intercommunicate solely by means of standardized message protocols;
    \item Easy implementation of complex agents through a full-featured hierarchy of base agent classes.
\end{itemize}
The focus on these features should enable an expanded range of experimental studies.  We believe ABIDES is also the first full-featured, modern market simulator to be shared with the community as an open source project.

\section{Important Questions Simulation Can Help Us Address}

ABIDES can support a number of different kinds of investigations into market behavior that are not easily conducted using historical data or live experiments.

\begin{itemize}
    \item \textbf{The benefits of co-location}: In the past 20 years hedge funds and other market participants have invested in the deployment of computing resources co-located at major exchanges \cite{zook2017}.  This so-called ``co-location'' enables quicker access to market information than if the trading server were located further away.  It is not feasible to investigate the value of the advantage co-location provides with available historical data, because it does not include information about  the geographic location, network latency, or network reliability of each actor. With a platform that does not require formal arms-length messaging using a realistic network model, we cannot simulate the effects of these factors even if they are known. ABIDES provides a network model and mandatory messaging protocol that enables detailed experiments in this area: Creating a population of agents with known distribution of network latency and reliability, conducting trials in which one agent is incrementally shifted from a co-location facility out to a great distance, and evaluating the impact of this shift on each agent's profitability while otherwise pursuing the same strategies.

\item \textbf{The impact of large orders on price}: The very act of trading, and even placing orders in a market may affect the price.  For instance, if there is significant selling pressure evidenced by a large volume of sell orders, it is generally expected that the price will go down.  The extent to which the price moves because of an order is referred to as \emph{market impact}.  Market participants of course want to minimize such impact, because the market usually moves contrary to their profit incentives.  In a market field study, it is not feasible to perform controlled A/B tests.  One cannot place a market buy at the NYSE for one million shares of IBM at 10 AM on Oct 22, 2018, and then also \emph{not} place that order, and compare the difference.  Without the ``control'', any observed result from the large order could be attributable to some other factor.  A key feature of ABIDES is the ability to re-simulate the same historical market day with known, limited changes while holding all other factors constant, thus enabling the desired experimental control population.

\item \textbf{Cost-benefit analysis of AI}: When analyzing historical market data, we cannot know the logic behind individual trader actions.  In a simulation without a model for computational time delays that directly impact time-to-market for the resulting orders, we cannot readily study the trade-off between simpler, faster predictors and slower, more powerful predictors.  ABIDES introduces a flexible, integrated model for computation delay that permits the ``speed'' of each agent's thought process to be represented, and to have that representation affect the timing of all of outbound messages as well as the next time at which the agent can be roused for participation.  Thus heavier thinkers will take longer to deliver a resulting order to the exchange and will be unable to act as frequently.

\item \textbf{Explanation of learning agent behavior}: A current key area of AI research across all application fields is explainability -- once taken for granted in classic knowledge-based AI, but now increasingly difficult with ``black box'' ML algorithms.  By providing a platform with centralized, time-synced event logging for all agents, we envision a ``clear box'' in which each agent's decision, intent, behavior, and result for every action are fully visible.  We hope to use this ability to dive deeply into the \emph{why} of learned policy functions.

\end{itemize}

\section{ABIDES Architecture} 

\begin{figure*}[h]
    \centering
    \includegraphics[width=6in]{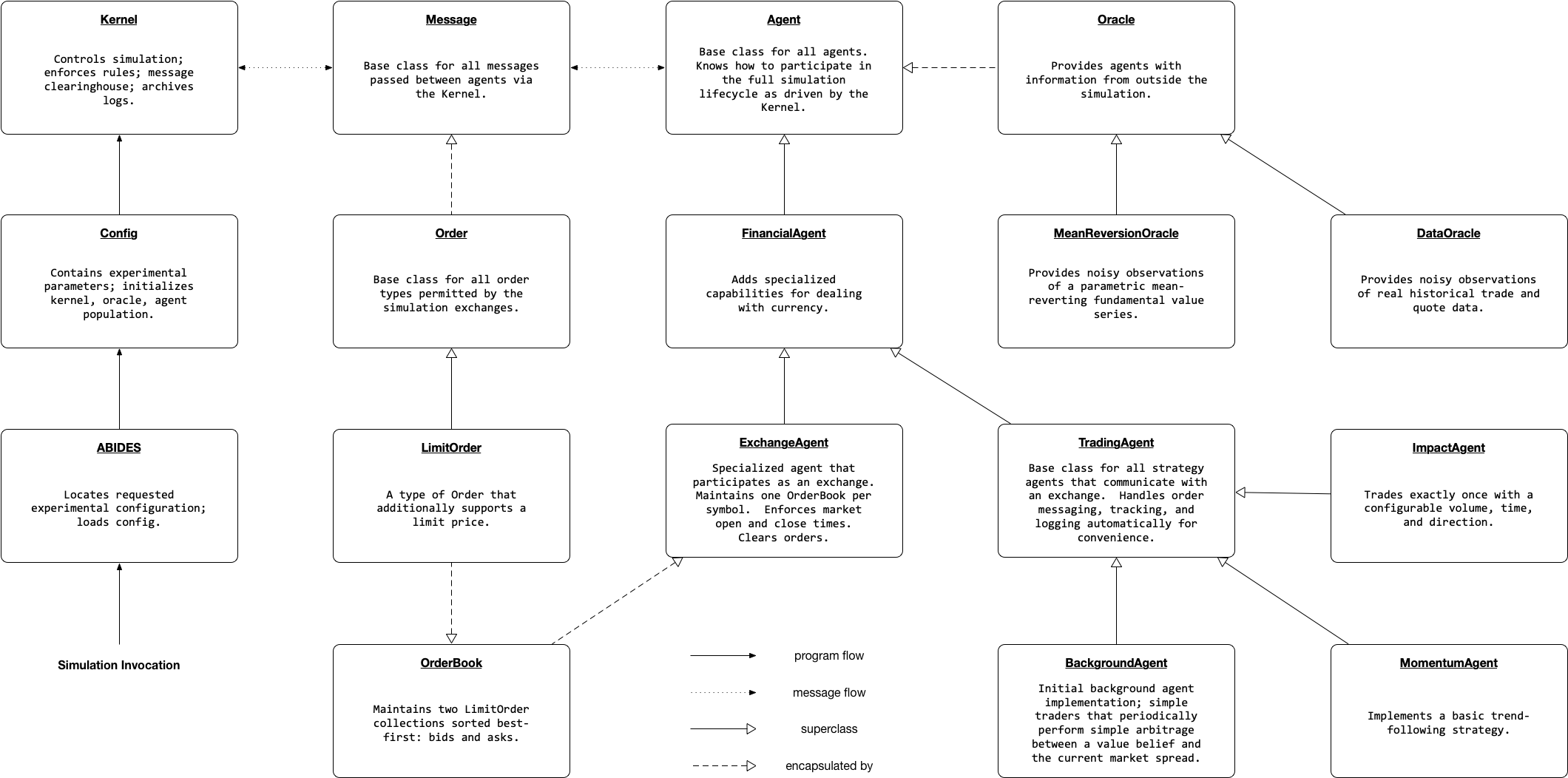}
    \caption{Class relations within the ABIDES simulation framework.}
    \label{fig:class_diagram}
\end{figure*}

The ABIDES framework includes a customizable configuration system, a simulation kernel, and a rich hierarchy of agent functionality as illustrated in Figure \ref{fig:class_diagram}.

\subsection{Functions and Features of the ABIDES Kernel}
The simulation is built around a discrete event-based kernel \cite{banks1998handbook}.   The kernel resides in the \texttt{Kernel} class in the default package and is required in all simulations.  All agent messages must pass through the kernel's event queue.  The kernel supports simulation of geography, computation time, and network latency.  It also acts as enforcer of simulation ``physics'', maintaining the current simulation time, tracking a separate ``current time'' for each agent, and ensuring there is no inappropriate time travel.  Key features of the ABIDES kernel include:

\begin{itemize}
\item \textbf{Historical date simulation} \hspace{1em} All simulation occurs on a configurable historical date or sequence of dates.  This permits ``real'' historical information to be seamlessly injected into the simulation at appropriate times when required for a particular experiment.

\item \textbf{Nanosecond resolution}: Because we seek to emulate real markets, we simulate time at the same resolution as an example exchange: the NASDAQ exchange. All simulation times are represented as Pandas Timestamp objects with nanosecond resolution.  This allows a mixture of agents to participate in the simulation on very different time scales with minimal developer overhead.  Events that occur simultaneously (in the same nanosecond) will be executed in arbitrary order.

\item \textbf{Global Virtual Time (GVT)}:   GVT is the latest simulated time for which all messages are guaranteed to have been processed.  The kernel tracks GVT as the simulation progresses.  It is usually the case that GVT advances much more quickly than wall clock time, but for very complex scenarios, it may not. The GVT value is not available to the agents.

\item \textbf{Current time per agent}: The kernel tracks a ``current time'' per individual participating agent which is incremented upon return from any call to \texttt{Agent.receiveMessage()} or \texttt{Agent.wakeup()}.  In situations where the current time for the agent is ``in the future'' (i.e., larger than GVT), the kernel will delay delivery of messages or wakeup calls to this agent until GVT catches up.

\item \textbf{Computation delay}: The kernel stores a computation delay per agent which is added to the agent's ``current time'' after each activity. The delay is also added to the sent time and delivery time of any outbound message from an agent to account for the agent's computation effort.  Agents may alter this computation delay to account for different sorts of computation events.

\item \textbf{Configurable network latency}: The kernel maintains a pairwise agent latency matrix and a latency noise model which are applied to all messages between agents.  This permits simulation of network conditions and agent location, including co-location.

\item \textbf{Deterministic but random execution}: The kernel accepts a single pseudo-random number generator (PRNG) seed at initialization.  This PRNG is then used to generate seeds for an individual PRNG object per agent, which must rely solely on that object for stochastic methods.  Since our system is single-threaded, this allows the entire simulation to be guaranteed identical when the same seed is initialized within the same experimental configuration.  This would not ordinarily permit the desired A/B testing, because the ``agent of change'' might consume an additional pseudo-random number from the sequence and thus change the stochastic source for all subsequent agents.  Because of our careful use of the primary PRNG only to generate subsidiary PRNGs per agent, the ``agent of change'' in an ABIDES A/B experiment will not alter the set of pseudo-random numbers given to any other agent throughout the simulation, even if it uses more or fewer such inputs for its changed activity.  In this way, changes in the behavior of other agents will be caused by a changed simulation environment (e.g. stock prices) and not simple stochastic perturbation.

\end{itemize}

\subsection{ABIDES Kernel Lifecycle Phases}
During a simulation, the kernel follows a series of life cycle phases.  All except the event queue processing phase consist entirely of sending the relevant event notification to all agents, and are described in the \textbf{Agent} subsection below.  The event queue processing phase is elaborated upon here:

\begin{enumerate}
    \item Kernel Initializing
    \item Kernel Starting
    \item Repeat until the event queue is empty or \texttt{currentTime} $>$ \texttt{stopTime} :
    \begin{itemize}
        \item[$-$] Extract next scheduled \texttt{event} and set \texttt{currentTime} $=$ \texttt{event.deliveryTime}
        \item[$-$] If \texttt{agentTimes[event.target]} $>$ \texttt{currentTime}:
        \begin{enumerate}
            \item[$\cdot$] \texttt{event.deliveryTime} $=$ \texttt{agentTimes[event.target]}
            \item[$\cdot$] Place event back in queue and \texttt{goto 3}
        \end{enumerate}
        \item[$-$] \texttt{agentTimes[event.target]} $=$ \texttt{event.deliveryTime}
        \item[$-$] Call \texttt{target.wakeup()} or \texttt{target.receiveMessage()}
        \item[$-$] \texttt{agentTimes[event.target] += computationDelay[event.target]}
    \end{itemize}
    \item Kernel Stopping
    \item Kernel Terminating
\end{enumerate}

The kernel additionally supports a few critical methods upon which agents depend:

\begin{itemize}
    \item \texttt{sendMessage(sender, recipient, message, delay)} - Schedules \texttt{message} to be transmitted from \texttt{sender} to \texttt{recipient} with an (optional) non-negative additional delay.  The ``sent time'' will be the sender's current time, plus its computation delay, plus any requested extra delay.  The ``delivery time'' will be the sent time plus network latency plus jitter, as determined by configured parameters for the experiment.
    \item \texttt{setWakeup(sender, requestedTime)} - Schedules a wakeup call for the sender at the requested future time.
    \item \texttt{findAgentByType(type)} - Returns the numeric identifier of an agent of the requested type if one can be found.  If multiple agents of the type exist, one is selected arbitrarily.  It is not possible for an agent to obtain a \emph{reference} to another agent (and thus bypass the kernel in the future).
    \item \texttt{writeLog(sender, dfLog)} - Called by an agent to request that its log be archived to disk for analysis.  The log is expected to be a Pandas DataFrame with index type DatetimeIndex.
\end{itemize}

\subsection{The Agent Class} All simulator agents are defined in the \texttt{agent} package.  All participants in a simulation must inherit from the base \texttt{agent.Agent} class, which implements a number of required methods that allow basic participation in the full life cycle of the simulation.

The following methods must be supported by all simulation agents and will be called exactly one time per agent by the kernel.  The order in which agents are activated in each life cycle phase is arbitrary.

\begin{itemize}
    \item \texttt{kernelInitializing(kernel)} - The kernel has just started running.  The existence of other agents should not be assumed.  There is no ``current time''.  The base \texttt{Agent} simply retains the given \texttt{kernel} reference.
    \item \texttt{kernelStarting(startTime)} - Event queue processing is about to begin.  All other agents are now guaranteed to exist.  There is no ``current time''.  \texttt{startTime} contains what will be the initial simulation timestamp.  The base \texttt{Agent} requests a wakeup call for this initial timestamp.
    \item \texttt{kernelStopping()} - Event queue processing has just ended.  All other agents are still guaranteed to exist.  There is no longer a ``current time''.  The base \texttt{Agent} takes no action.
    \item \texttt{kernelTerminating()} - The kernel is about to shut down.  The existence of other agents should not be assumed.  There is no longer a ``current time''.  Agents are expected to log any final data and clean up.  The base \texttt{Agent} passes off its individual event log, if there are entries, to the kernel for archival.
\end{itemize}

The following methods must be supported by all simulation agents.  They will be called by the kernel in order of increasing delivery timestamp of queued messages and wakeup calls.  In both cases, the base \texttt{Agent} simply updates its internal current time and displays an informative message.

\begin{itemize}
    \item \texttt{receiveMessage(currentTime, msg)} - The kernel is delivering a message from another agent.  \texttt{currentTime} is the current simulation time as a Pandas Timestamp (nanosecond resolution).  \texttt{msg} is an instance of class \texttt{message.Message} which the agent must interpret.
    \item \texttt{wakeup(currentTime)} - The kernel is delivering a previously-scheduled ``wakeup call'' to the agent.  \texttt{currentTime} is the current simulation time.  No message is delivered, thus the agent must use internal state and logic to determine what it should do next.
\end{itemize}

While not required by the simulation kernel, the base \texttt{Agent} class also provides \texttt{logEvent(eventType, event)}, which can be called by any agent to append to an individual timestamped log of events.  As noted above, by default this log is reported to the kernel for archival during the \texttt{kernelTerminating} life cycle phase.

\subsection{The Exchange Agent Class}

The \texttt{agent.ExchangeAgent} class inherits from \texttt{agent.Agent} and represents a stock exchange such as NASDAQ.  The message protocols supported by this agent are based on NASDAQ's published ITCH and OUCH protocols. \cite{itch,ouch} The exchange is initialized with market opening and closing times, which it will enforce.  These are not required to match the simulation start and stop times.  The exchange agent is not privileged in any way; it must participate in the simulation just as any other agent. The \texttt{ExchangeAgent} understands how to respond to these types of messages:

\begin{itemize}
\item \textbf{Market Open Time}: Returns the timestamp at which the exchange will begin processing order-related messages.

\item \textbf{Market Close Time}: Returns the timestamp at which the exchange will stop processing order-related messages.

\item \textbf{Query Last Trade}: Returns the last trade price for a requested symbol.  Until the first trade of the day, the exchange reports the oracle open price (historical or generated data) as the ``last trade price''.  The exchange does not yet implement the opening cross auction.

\item \textbf{Query Spread / Depth}: Returns a list of the $N$ best bid and best ask prices for a requested symbol and the aggregate volume available at each price point.  With a requested depth of one, this is equivalent to querying ``the spread''.

\item \textbf{Limit Order}: Forwards the attached limit order to the requested symbol's order book for matching or acceptance.  Agents currently simulate market orders using a limit order with an arbitrarily high or low limit price.

\item \textbf{Cancel Order}: Forwards the attached order to the requested symbol's order book to attempt cancellation.
\end{itemize}

Outside of market hours, the exchange will only honor messages relating to market hour inquiries and final trade prices (after the close).  The exchange sends a ``market closed'' message to any agent which contacts it with disallowed messages outside of market hours.

\begin{figure*}[h]
    \centering
    \includegraphics[width=6in]{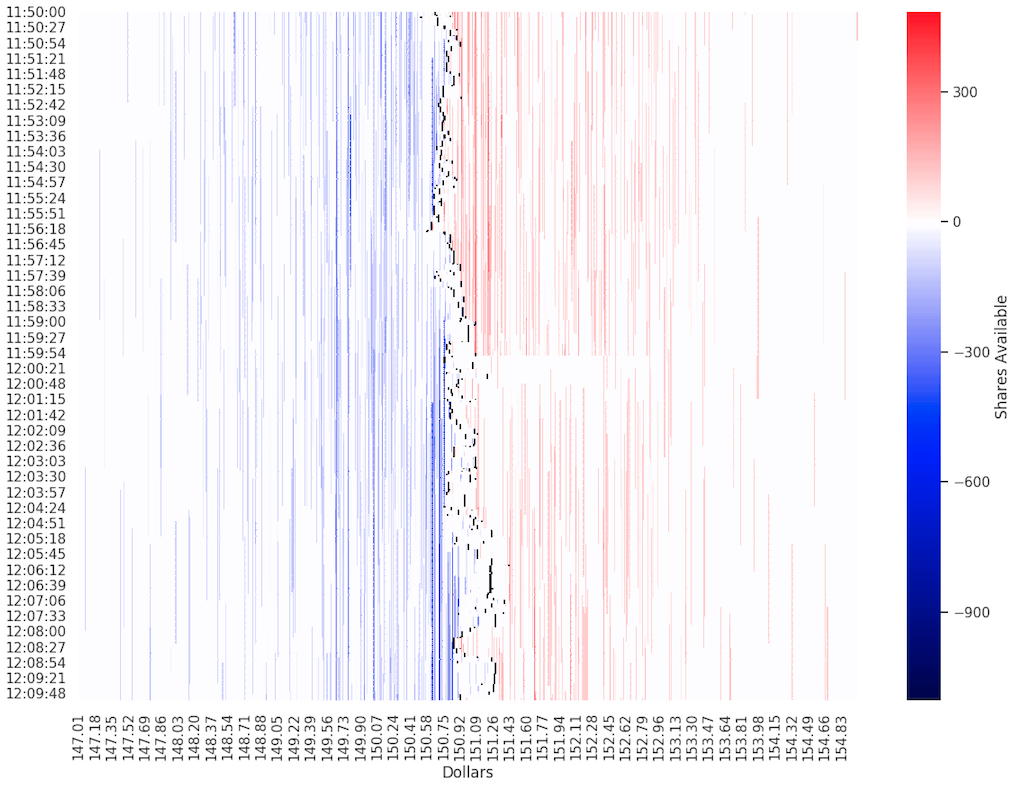}
    \caption{Example of order book visualization around the time of a high impact trade.}
    \label{fig:order_book}
\end{figure*}

The exchange agent demonstrates one use of the inbuilt Kernel logging facility, recording either the full order stream or snapshots of its order books at a requested frequency, enabling extremely detailed visualization and analysis of the order book at any time during simulation.  For example, Figure \ref{fig:order_book} shows a time window surrounding one ``high impact'' market buy order, which drives prices upward immediately and has a follow-on effect on other agents' value beliefs.

\subsection{The Order Book}

Within an Exchange Agent, an order book tracks all open orders, plus the last trade price, for a single stock symbol.  All order book activity is logged through the exchange agent.  The order book implements the following functionality:

\begin{itemize}

\item \textbf{Order Matching} Attempts to match the incoming order against the appropriate side of the order book.  The best price match is selected.  In the case of multiple orders at the same price, the oldest order is selected.

\item \textbf{Partial Execution} Either the incoming order or the matched limit order may be partially executed.  When the matched limit order is partially executed, the order is left in the book with its quantity reduced.  When the incoming order is partially executed, its quantity is reduced and a new round of matching begins.  Participants receive one ``order executed'' message, sent via the exchange, per partial execution noting the fill price of each.  When the incoming order is executed in multiple parts, the average price per share is recorded as the last trade price for the symbol.

 \item \textbf{Order Acceptance} When the incoming limit order has remaining quantity after all possible matches have been executed, it will be added to the order book for later fulfillment, and an ``order accepted'' message will be sent via the exchange.

\item \textbf{Order Cancellation} The order book locates the requested order by unique order id, removes any remaining unfilled quantity from the order book, and sends an ``order cancelled`` message via the exchange.

\end{itemize}

One might reasonably expect the order book in a market simulation to include a model for slippage.  We assert that our platform produces realistic slippage naturally, without the need for such a model.  Orders directed to the exchange suffer dynamic computation and network delays, during which time other orders are being executed.

\subsection{The Trading Agent Class}

The \texttt{agent.TradingAgent} class inherits from \texttt{agent.Agent} and represents the base class for a financial trading agent.  It implements a number of additional features beyond the basic simulator \texttt{Agent}, upon which subclassed strategy agents may rely:

\begin{itemize}

\item \textbf{Portfolio} The base trading agent maintains an equity portfolio including a cash position.  It automatically updates this portfolio in response to ``order executed'' messages.

\item \textbf{Open Orders} The trading agent keeps a list of unfilled orders that is automatically updated upon receipt of ``order executed'' and ``order cancelled'' messages, and when new orders are originated.

\item \textbf{Last Known Symbol Info} The trading agent tracks known information about all symbols in its awareness, including the most recent trade prices, daily close prices (after the close), and order book spread or depth.  These are automatically updated when receiving related messages.

\item \textbf{Market Status}  Upon initially waking at simulation start, the trading agent automatically locates an exchange agent, requests market open and close times, and schedules a second wakeup call for the time of market open.  It also maintains and provides a simple ``market closed'' flag for the benefit of subclassing agents.

\item \textbf{Mark to Market} The trading agent understands how to mark its portfolio to market at any time, using its most current knowledge of equity pricing.  It automatically marks to market at the end of the day.

\item \textbf{Messages} The trading agent knows how to originate all of the messages the exchange understands, and to usefully interpret and store all of the possible responses from the exchange.

\item \textbf{Logging} The trading agent logs all significant activity: when it places orders; receives notification of order acceptance, execution, or cancellation; when its holdings change for any reason; or when it marks to market at the end of the day.
\end{itemize}

\section{ABIDES Implementation}

The ABIDES simulator is implemented using Python, currently 3.6,  and the data analytical libraries NumPy \cite{oliphant2006guide}, and Pandas \cite{mckinney2010data}.  It makes use of a virtual environment to provide platform independence and provides a straightforward deployment.  It is seamlessly built to facilitate quick reconfiguration of varying agent populations, market conditions, exchange rules, and agent hyperparameters.  

Basic execution of the simulation can be as simple as: \texttt{python abides.py -c config}, where \texttt{config} is the name of an experimental configuration file.  Additional command line parameters are forwarded to the configuration code for processing, so each experimental configuration can add its own required parameters to a standard interface.  Complex experimental configuration can be performed directly within the config file since it is simply Python code, however the inclusion of command line arguments is beneficial for coarse grain parallelization of multiple experiments of the same type, but with varied simulation parameters.

A typical configuration file will specify a historical date to simulate and a simulation start and stop time as a nanosecond-precision \texttt{pandas.Timestamp} objects.   It will then initialize a population of agents for the experiment, configuring each as desired.  For example, an experiment could involve 1,000 background agents (perhaps Zero Intelligence agents or Heuristic Belief Learning agents), 100 high-frequency trading agents, and one impact agent with various initialization parameters to control their behavior.  Each agent will at least be given a unique identifier and name.  The configuration file will also construct a latency matrix (pairwise between all agents at nanosecond precision) and latency noise model which will be applied to all inter-agent communications.  If a ``data oracle'', a utility with access to a data source outside the simulation, is required for the experiment, the configuration file will initialize one.  Finally a simulation kernel will be initialized and run, passing it the agent population, oracle, and other simulation parameters.

Note that there is nothing finance-specific about the bootstrapper, configuration template, simulation kernel, or the base \texttt{Agent} class.  All are appropriate for use in any continuous-time discrete event simulation.

\subsection{Example: A Momentum Trading Agent}

To highlight the simplicity of creating a functional trading agent in our simulated environment, we present the code for a basic momentum trader.  It wakes each minute during the day, queries the last trade price, projects a future price using linear regression over a configurable last $N$ data points, and places a market order based on this projection.  Following is the complete source, excluding import statements:

\begin{lstlisting}
class MomentumAgent(TradingAgent):

  def __init__(self, id, name, symbol, startingCash, lookback):
    super().__init__(id, name, startingCash)

    self.symbol = symbol
    self.lookback = lookback
    self.state = "AWAITING_WAKEUP"

    self.trades = []


  def wakeup (self, currentTime):
    can_trade = super().wakeup(currentTime)

    if not can_trade: return

    self.getLastTrade(self.symbol)
    self.state = "AWAITING_LAST_TRADE"


  def receiveMessage (self, currentTime, msg):
    super().receiveMessage(currentTime, msg)

    if self.state == "AWAITING_LAST_TRADE" and \
         msg.type == "QUERY_LAST_TRADE":
         
      last = self.last_trade[self.symbol]
      self.trades = (self.trades + [last])[:self.lookback]

      if len(self.trades) >= self.lookback:
        m, b = np.polyfit(range(len(self.trades)), self.trades, 1)
        pred = self.lookback * m + b

        holdings = self.getHoldings(self.symbol)

        if pred > last:
          self.placeLimitOrder(self.symbol, 100-holdings,
                               True, self.MKT_BUY)
        else:
          self.placeLimitOrder(self.symbol, 100+holdings,
                               False, self.MKT_SELL)

      self.setWakeup(currentTime + pd.Timedelta("1m"))
      self.state = "AWAITING_WAKEUP"
\end{lstlisting}

\section{Case Study: Background Agents}

A long-term goal is to produce realistic but possibly noisy re-simulations of particular days in history to play out various ``what if'' scenarios. The idea is to populate the simulation with a large number of trading agents that provide a realistic environment into which experimental agents can be injected.

\begin{figure*}[h]
    \centering
    \subfloat[IBM: September 30, 2008]{
        \includegraphics[width=2.5in]{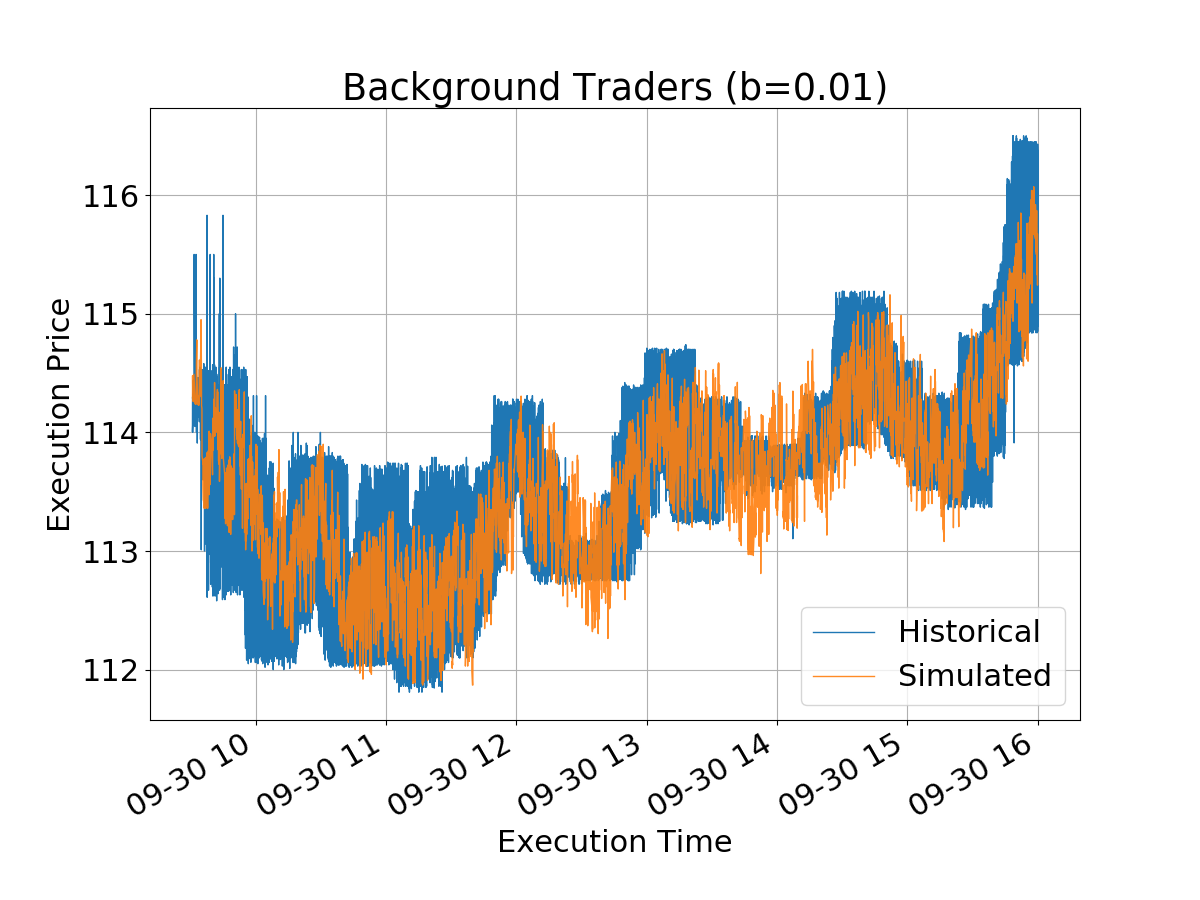}
    }
    \subfloat[MSFT: June 24, 2016]{
        \includegraphics[width=2.5in]{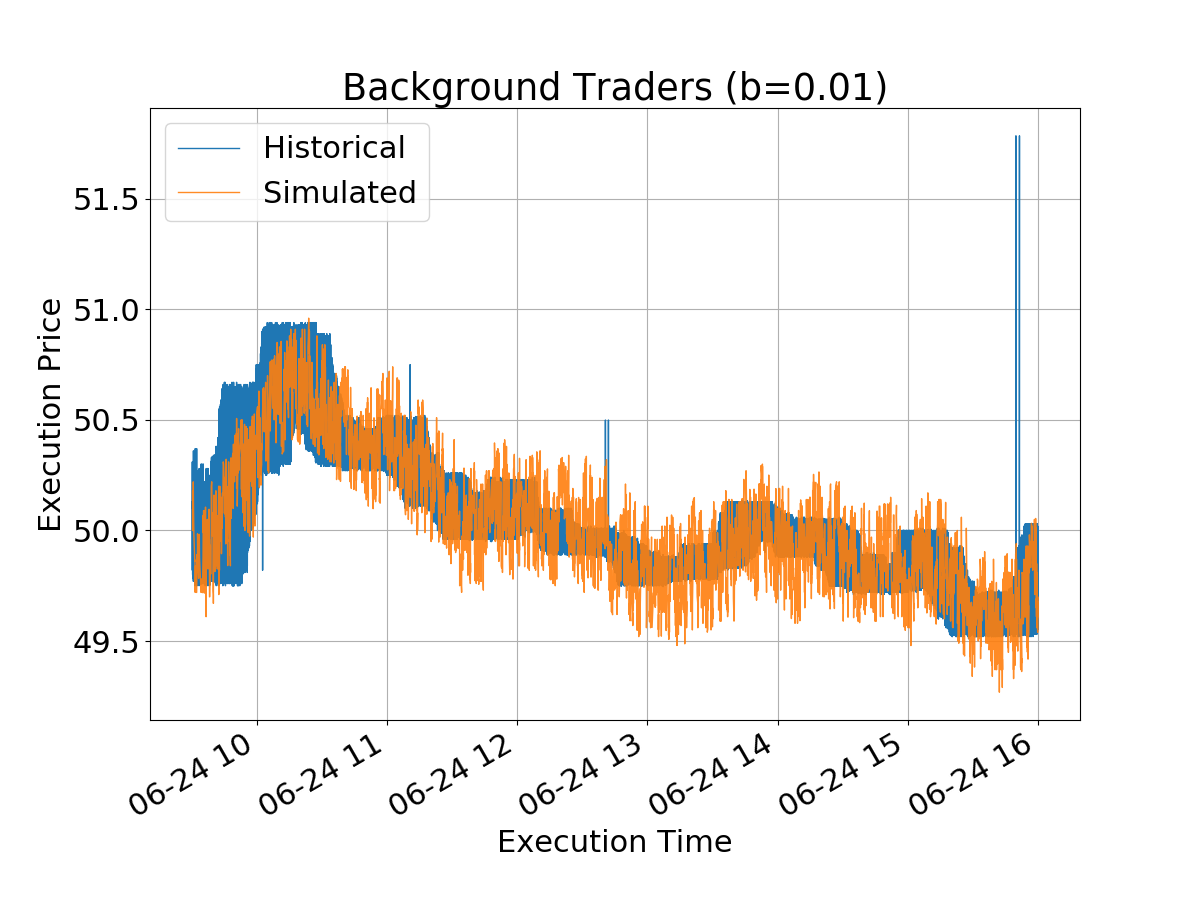}
    }
    \caption{Simulated trades versus historical trades on two days.}
    \label{fig:background}
\end{figure*}

Our initial effort towards this goal involves the introduction of a data oracle with access to fine-resolution historical trade information, and the creation of a ``background'' agent which is able to request a noisy observation of the most recent historical trade as of the agent's current simulated time. The approach is meant to reproduce the behavior of a trader whose beliefs regarding the fundamental value of a stock are informed by interpretations of news and other incoming information.  It was inspired by the concept of a stock's ``fundamental value'' as used in the work of Wang and Wellman. \cite{wang2017spoofing}  Our approach is similar, but it uses historical data as a baseline rather than a mean-reverting stochastic process.

A common baseline agent in the continuous double auction literature is the Zero Intelligence (ZI) trader \cite{gode1993allocative} which submits random bids and offers to the market, usually drawn from some stochastic distribution around a central value belief for the underlying instrument.  

Our \texttt{agent.BackgroundAgent} class follows the general spirit of the ZI trader, but with two important distinctions: The central value belief at any time is a mixture of the prior belief with a noisy observation of a historical trade; and the agent implements an extremely basic arbitrage strategy between the last simulated trade price and its internal belief.  Thus the valuation is influenced by random factors, but the direction of limit orders placed is then rational, with the agent assuming the simulated price will converge to its value belief over time.  Each background agent trades only a single symbol on a single exchange.

In our current configuration, a background agent typically follows the following basic logic, given some wake frequency $F$ in some unit of time (microseconds, seconds, etc):

\begin{enumerate}
    \item Request an initial wakeup time selected randomly from a uniform distribution across the first $F$ interval after market open
    
    \item On wakeup, cancel any unfilled orders and wait for confirmation.
    
    \item Query the exchange for the last trade price of this agent's symbol of interest and wait for the response.
    
    \item Request a new noisy historical observation from the data oracle, and mix this observation with any prior belief to obtain a new posterior value belief.
    
    \item Determine the direction from the simulated last trade to this agent's value belief.  Place a limit order to bring the agent's holdings in line with a presumed profitable position: entering, exiting, or reversing position as necessary.
    
    \item Request a new wakeup call for the current time plus approximately $F$.
\end{enumerate}

Figure \ref{fig:background} compares the behavior of 100 background agents interacting in ABIDES with the actual intra-day price on two separate days in history.  Ideally, we will see a price history that closely resembles the day in history, with similar statistical properties.

\section{Case Study: Market Impact}

\begin{figure*}[h]
    \centering
    \includegraphics[width=6in]{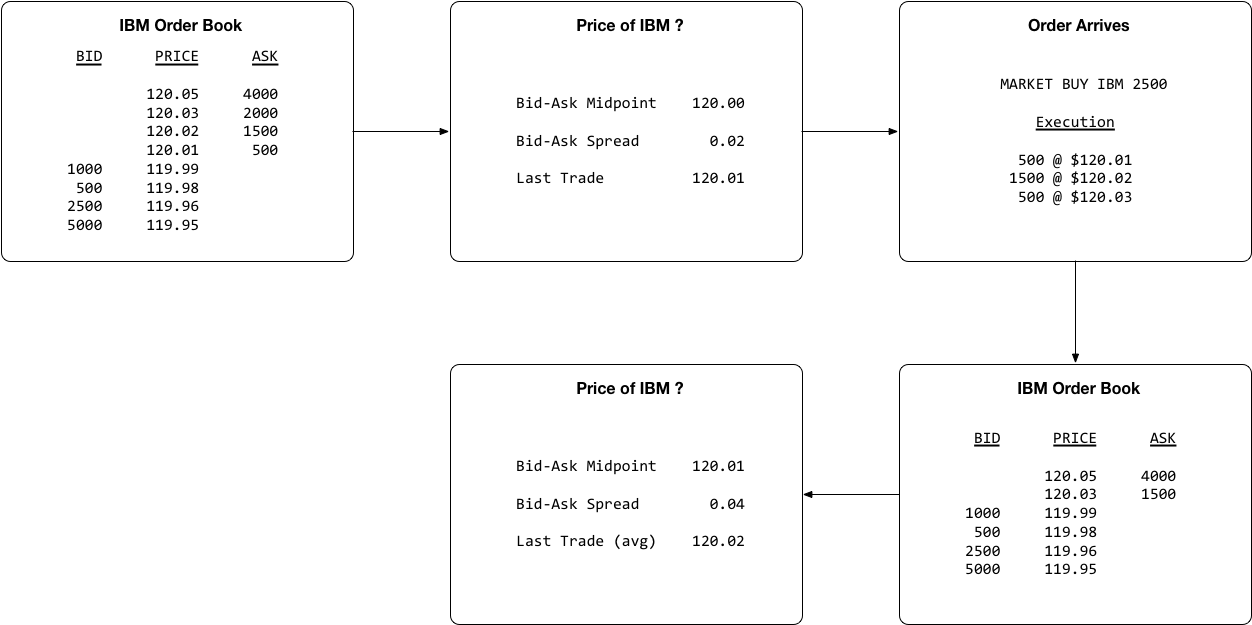}
    \caption{Example of mechanical market impact.}
    \label{fig:impact_example}
\end{figure*}

One area in which we believe simulation can add significant value to the current state of knowledge in finance is more accurate models of the market impact of large trades.  Each order placed at the exchange potentially ``moves the market'' due to the nature of the market microstructure within the order book: arriving orders can add liquidity at a better price, altering the spread; or can match existing orders and remove liquidity from the market.  See Figure \ref{fig:impact_example} for an example of mechanical market impact.

Models that rely on historical data encounter limitations stemming from the inability to repeat history while introducing an experimental change and allowing subsequent events to be \emph{altered} by that change.  Models can attempt to compare ``similar'' days in history, but no two market days are ever the same.

If one could instead create a multi-agent simulation of a particular date in history such that a near approximation of historical trades emerged in the absence of any significant change, but the trading agents would realistically react to any such changes, a more accurate understanding of large trade impact could be attained. Here we present a preliminary investigation of this idea.

We begin each simulation with a population of background agents and at least one exchange agent.  For this experiment, we add a single experimental agent, \texttt{agent.ImpactAgent}, which simply places a single large market order at a predetermined time of day.  The experimental parameter for the agent is its ``greed''; that is, the proportion of available order book liquidity near the spread it consumes at the time of trade.  For example, a long impact agent with $greed=0.1$ will place a market buy order for $10\%$ of the shares on offer.

\begin{figure*}[h]
    \centering
    \subfloat[MARKET BUY 1232 IBM]{
        \includegraphics[width=2.5in]{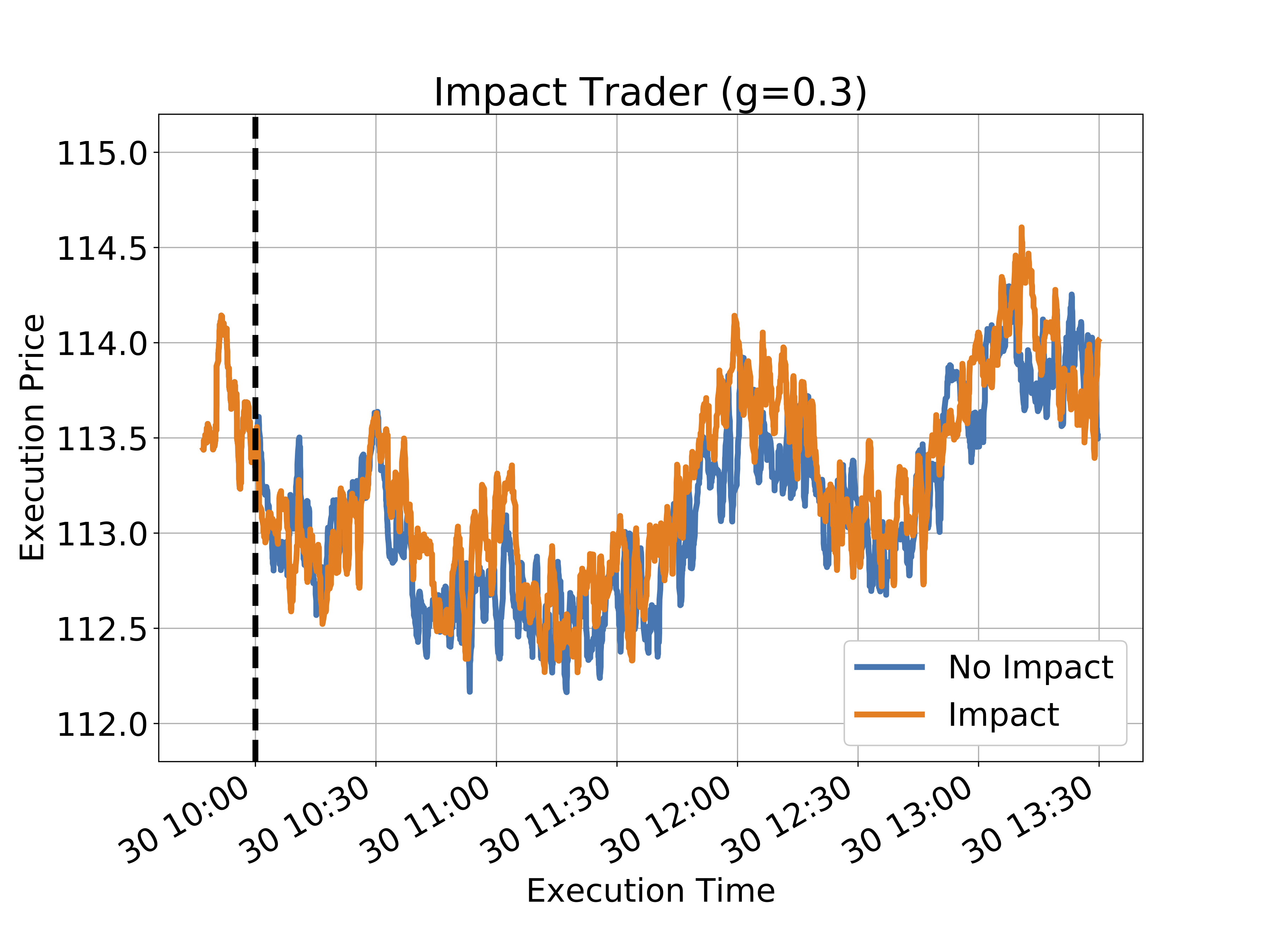}
    }
    \subfloat[MARKET BUY 2874 IBM]{
        \includegraphics[width=2.5in]{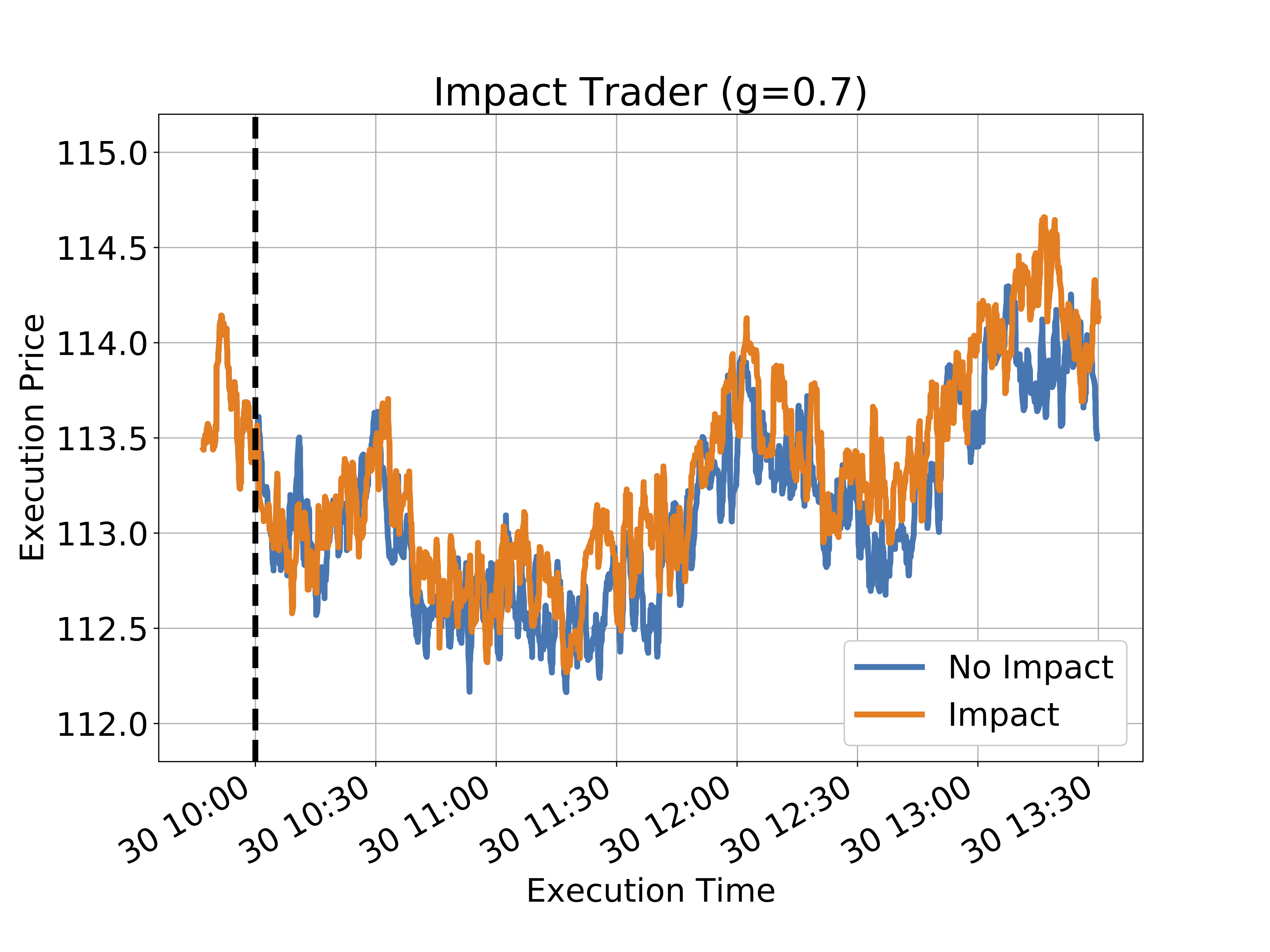}
    }
    
    \subfloat[MARKET BUY 5338 IBM]{
        \includegraphics[width=2.5in]{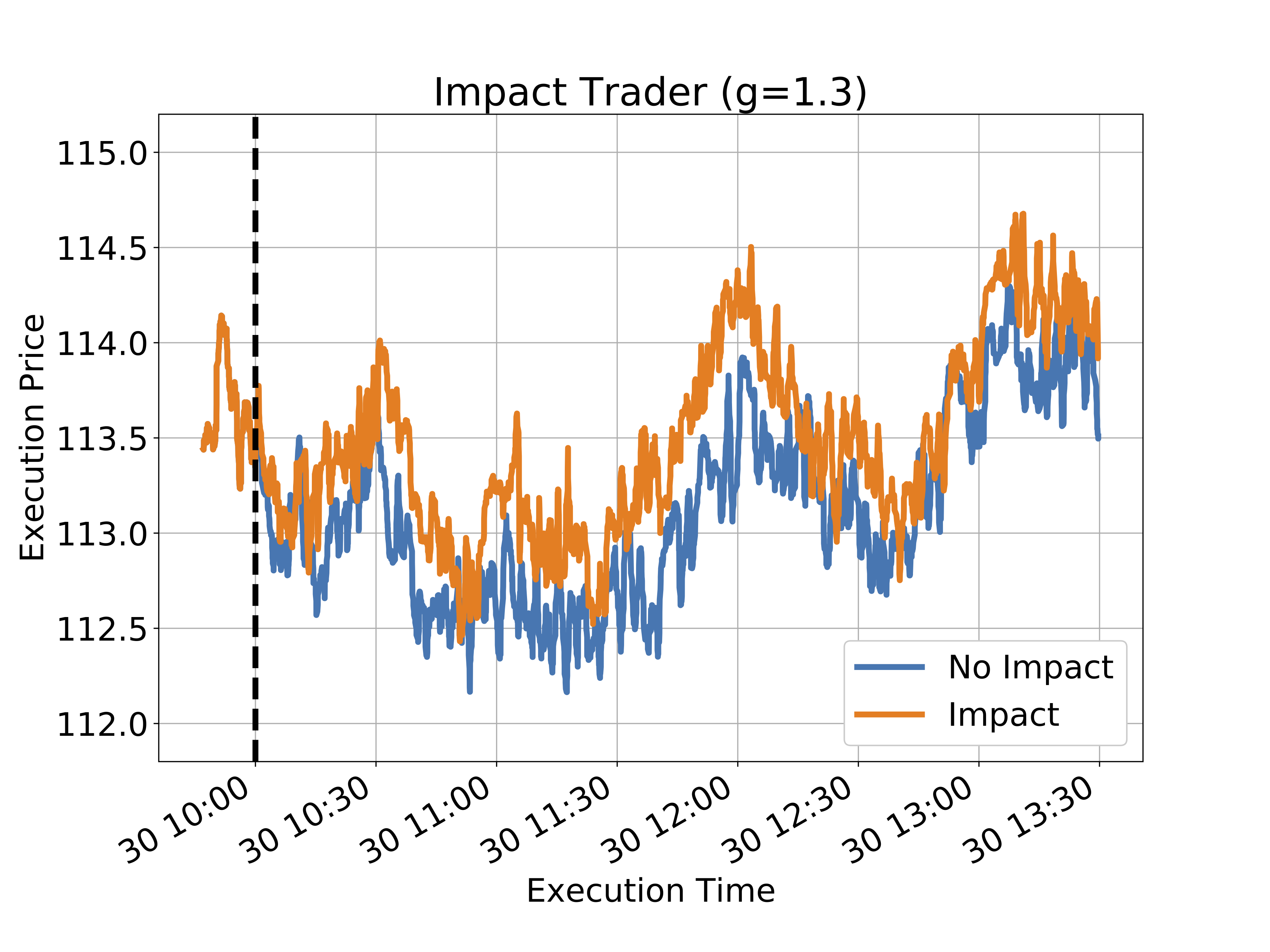}
    }
    \subfloat[MARKET BUY 7801 IBM]{
        \includegraphics[width=2.5in]{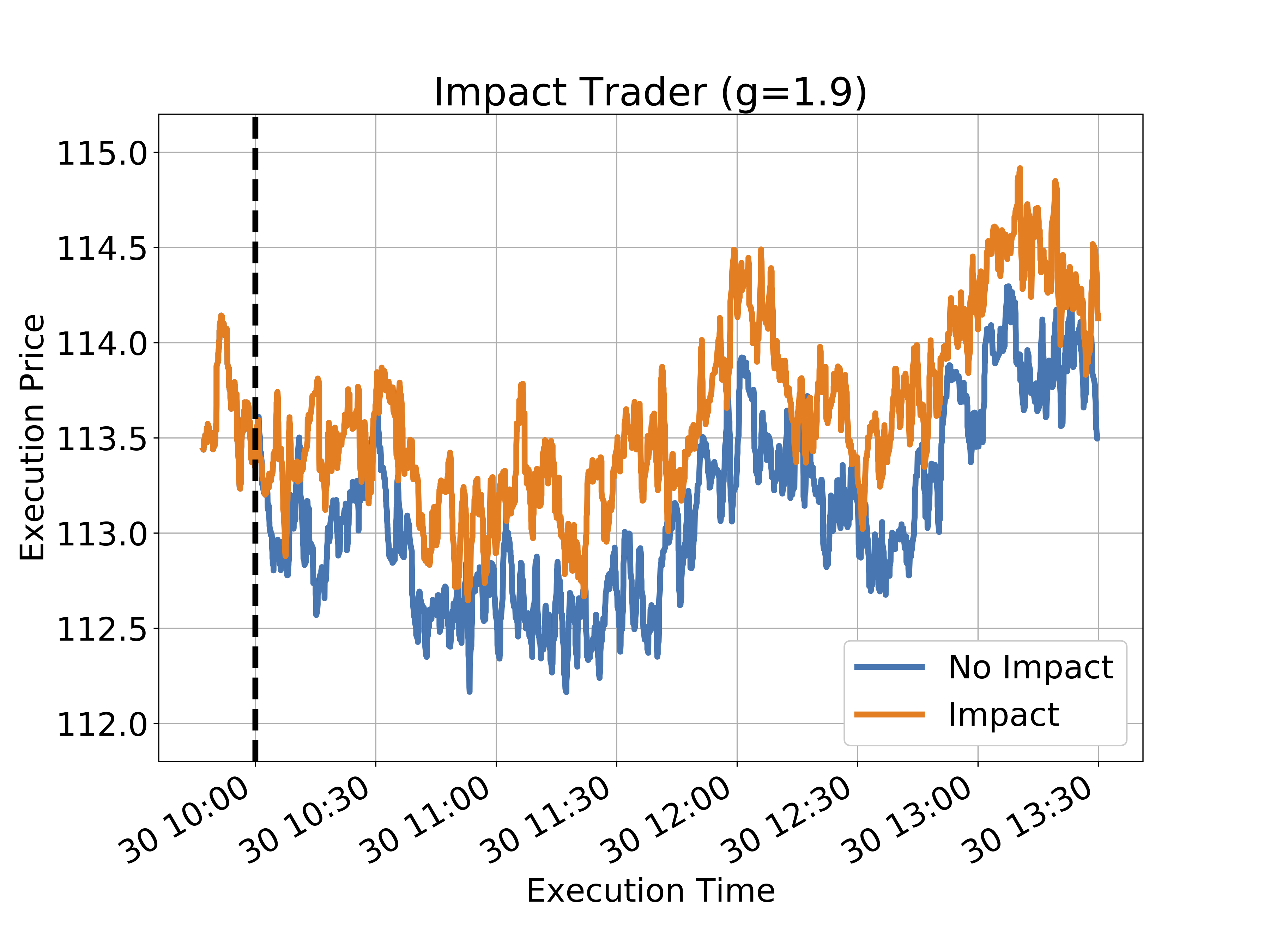}
    }
    \caption{Market impact of trades at 10:00 AM.}
    \label{fig:impact}
\end{figure*}

Our experiment includes 100 background agents and one exchange agent handling an order book for a set of symbols including IBM.  In Figure \ref{fig:impact}, the blue line represents each trade made by our population of background agents in the absence of an impact trader.  The orange line shows each trade made by the simulated trading agents given the introduction of a single impact agent with varying ``greed'', acting one time with one trade at 10:00 AM on September 30, 2008.  Both series are smoothed to improve visibility of the differences.

\begin{figure*}[h]
    \centering
    \subfloat[Impact agent with greed 0.5]{
        \includegraphics[width=2.5in]{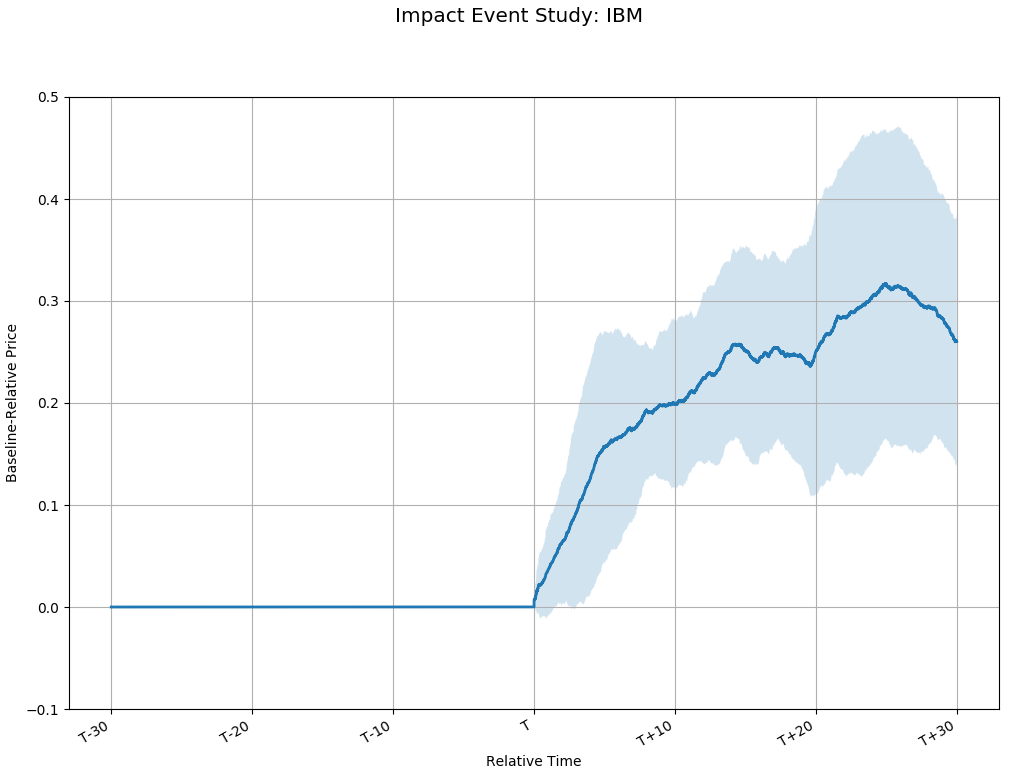}
    }
    \subfloat[Impact agent with greed 0.1]{
        \includegraphics[width=2.5in]{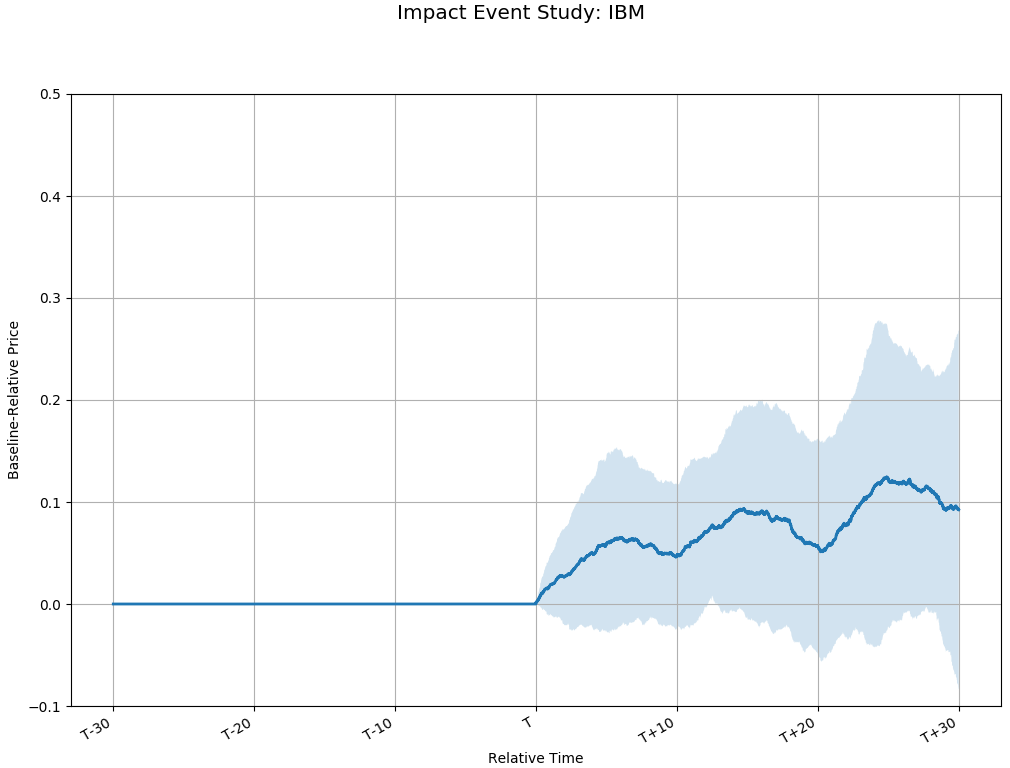}
    }
    \caption{Market impact event studies.}
    \label{fig:event}
\end{figure*}

The impact trader has a clear effect on the market, despite the background agents' central tendency to arbitrage the price toward historical levels, and the impact grows larger proportionally with its market bid size.  The change is particularly noticeable in the cyclical peaks of the auction.  Due to the price elevation it caused, the impact trader's total profit increased with the size of its bid from an average of \$2,633 with $greed=0.3$ to \$12,502 with $greed=1.9$.  However its profit per share declined from \$2.14 to \$1.60.  We found a correlation between profit per share and trade size of $r=-0.31$ across sixty experimental trials.

It is useful to consider these market impacts in aggregate across multiple experimental examples.
ABIDES makes it easy to produce study plots from logged simulation data.  Figure \ref{fig:event} shows a time-aligned event study of many impact trades at different times, on different days, to illustrate the range of likely price effects after the time of impact.

\section{Conclusion and Future Challenges}

We presented the design and implementation of ABIDES, a high-fidelity equity market simulator. ABIDES provides an environment within which complex research questions regarding trading agents and market behavior can be investigated. 

The simulation is demonstrated in two case studies. The first case study shows how previous intra-day transaction histories are closely reproduced by a population of interacting background trading agents communicating with an exchange agent.  These background agents are designed to provide a realistic market environment into which experimental agents can be injected.  The second case study illustrates how large market orders impact simulated prices not just immediately, but for a significant period after the order arrives at the exchange.  It is also intended to demonstrate the experimental potential of the ABIDES platform.

We now have a robust simulation environment in which to develop and experiment with more complex trading agents, including those based on approaches in machine learning and artificial intelligence.

\section{Open Source Access and License}

ABIDES is available through GitHub at \texttt{https://github.com/abides-sim/abides} under the BSD 3-clause license.

\section{Acknowledgements}

This material is based upon research supported by the National Science Foundation under Grant No. 1741026.

\newpage

\begin{quote}
\begin{small}
\bibliographystyle{acm}
\bibliography{article}
\end{small}
\end{quote}

\end{document}